\documentclass[12pt]{iopart}

\begin{document}

\title[Natural Cutoffs and Quantum Tunneling from Black Hole Horizon]
{Natural Cutoffs and Quantum Tunneling from Black Hole Horizon}

\author{{ Kourosh Nozari \footnote{knozari@umz.ac.ir}}\quad and\quad  { Sara Saghafi  \footnote{s.saghafi@stu.umz.ac.ir}}}

\address{Department of Physics, Faculty of Basic Sciences,\\ University of Mazandaran, P. O. Box 47416-95447,\\ Babolsar, IRAN}

\begin{abstract}

We study tunneling of massless particles through quantum horizon of
a Schwarzschild black hole where quantum gravity effects are taken
into account. These effects are encoded in the existence of natural
cutoffs as a minimal length, a minimal momentum and a maximal
momentum through a generalized uncertainty principle. We study
possible correlations between emitted particles to address the
information loss problem. We focus also on the role played by these
natural cutoffs on the tunneling rate through quantum horizon.
\end{abstract}
\maketitle

\section{Introduction}\vspace{2mm}

Incorporation of gravity in quantum field theory leads naturally to
an effective cutoff; a minimal measurable length in the ultraviolet
regime. In fact, the high energies used to probe small distances
significantly disturb the spacetime structure by their powerful
gravitational effects. Some approaches to quantum gravity proposal
such as string theory \cite{1}-\cite{7}, loop quantum gravity
\cite{8} and quantum geometry \cite{9} all indicate the existence of
a minimal measurable length of the order of the Planck length,
$\ell_{p}\sim 10^{-35}$m (see also \cite{10}-\cite{13}). Moreover,
some Gedanken experiments in the spirit of black hole physics have
also supported the idea of existence of a minimal measurable length
\cite{14}. So, the existence of a minimal observable length is a
common feature of all promising quantum gravity candidates. The
existence of a minimal measurable length modifies the Heisenberg
uncertainty principle (HUP) to the so called Generalized
(Gravitational) uncertainty principle (GUP) represented for instance
as $$\Delta x\Delta p\geq\hbar\Big[1+\alpha^2\ell_{p}^2(\Delta
p)^2\Big].$$

In the HUP framework there is essentially no restriction on the
measurement precision of particles' position, so that $\Delta x_{0}$
as the minimal position uncertainty could be made arbitrarily small
toward zero. But, this is not essentially the case in the GUP
framework due to existence of a minimal uncertainty in position
measurement. In this respect, there is a finite resolution of
spacetime points in quantum gravity regime.

On the other hand, Doubly Special Relativity theories (for review
see for instance \cite{15}) suggest that existence of a minimal
measurable length would restrict a test particle's momentum
uncertainty to take any arbitrary values leading nontrivially to an
upper bound, $P_{{{\it max}}}$, on a test particle's momentum. This
means that there is a maximal particle's momentum due to the
fundamental structure of spacetime at the Planck scale
\cite{16}-\cite{19}. Such a smallest length and maximal momentum
cause an alteration of the position-momentum uncertainty relation in
such a way that
\begin{equation}
\Delta x\Delta p\geq\hbar\Big[1-\alpha\ell_{p}(\Delta
p)+\alpha^2\ell_{p}^2(\Delta p)^2\Big]
\end{equation}
where $\alpha$ is a dimensionless constant of the order of unity
that depends on the details of the quantum gravity hypothesis. In
the standard limit, $\Delta x\gg\ell_{p}$, it yields the standard
Heisenberg uncertainty principle, $\Delta x\Delta p\geq \hbar$.

It has been shown that for a spherical black hole the thermodynamic
quantities can be obtained by using the standard uncertainty
principle \cite{20}. In the same manner, incorporation of quantum
gravity effects in black hole physics and thermodynamics through a
GUP with the mentioned natural cutoffs brings several new and
interesting implications and modifies the result dramatically.
Specially, the final stages of black hole evaporation obtains a very
rich phenomenology in this framework. Application of GUP with
minimal length to black hole physics has been widely studied in
recent years (see for instance \cite{21}-\cite{28}). In an elegant
work, Parikh and Wilczek constructed a procedure to describe the
Hawking radiation emitted from a Schwarzschild black hole as a
tunneling through its quantum horizon \cite{29}. This procedure
provides a leading correction to the tunneling probability (emission
rate) arising from the reduction of the black hole mass because of
the energy carried by the emitted quantum. However, because of the
lack of correlation between different emitted modes in the black
hole radiation spectrum, the form of the correction is not adequate
by itself to recover information. The Parikh-Wilczek tunneling in
the presence of a minimal measurable length and possible resolution
of the information loss problem in this framework has been studied
recently in \cite{24}(see also \cite{25} for treating the problem in
noncommutative framework). Here we generalize this work in two
steps: firstly we consider a GUP that admits both a minimal
measurable length and a maximal momentum (see \cite{30} and
references therein) and secondly the case that there are a minimal
length, a minimal momentum and a maximal momentum (see \cite{30} and
also \cite{31}). The later case is the most general treatment of the
problem with incorporation of all natural cutoffs. The crucial
difference of our study in this paper with our previous study
reported in \cite{24} is in the fact that \cite{24} ignores the
existence of a minimal and maximal momentum for emitted particles.
As we shall see, the minimal and maximal momentum arise respectively
by addition of terms proportional to $(\Delta x)^{2}$ and $\Delta p$
in the right hand side of relation (1). Existence of a minimal and a
maximal momentum for emitted particles have important implications
on the tunneling rate of particles through quantum horizon. We show
that in this framework, correlations between different modes of
radiation evolve, which reflects the fact that information emerges
continuously during the evaporation process at the quantum gravity
level. This feature has the potential to answer some questions
regarding the black hole information loss problem and provides a
more realistic background for treating the black hole evaporation in
its final stage of evaporation.

With these motivations, we apply back-reaction and quantum gravity
effects to study black hole evaporation process and thermodynamics.
We use suitable GUPs in each case to drive a modified black hole
temperature. Then, by using the first law of black hole
thermodynamics, we obtain modified black hole entropy. We show that
in the presence of new natural cutoffs as minimal and maximal
momentum, in the final stage of the black hole evaporation, when the
black hole mass is of the order of the Planck mass, the temperature
is more than the cases that we consider just the minimal length or
minimal length-maximal momentum schemes. We also show that the GUP
prevents black holes to evaporate totally, resulting a stable
remnant.

\section{Minimal Length, Maximal Momentum and the Parikh-Wilczek Tunneling Mechanism}\vspace{2mm}

The first step to discuss the quantum tunneling through the black
hole horizon is to find a proper coordinate system for the black
hole metric where all the constant lines are flat, and the tunneling
path is free of singularities. Painlev$\acute{e}$ coordinates are
suitable choices in this respect. In these coordinates, the
schwarzschild metric is given by
\begin{equation}
ds^2=-\Big(1-\frac{2m}{r}\Big)dt^2+\Big(2\sqrt{\frac{2m}{r}}\Big)dtdr+dr^2+r^2
 \Big(d\theta^2+sin^2\theta d\phi^2\Big)\,,
\end{equation}
which is stationary, non-static, and non-singular at the horizon.
The radial null geodesics are given by
\begin{equation}
\dot{r}\equiv\frac{dr}{dt}=\pm1-\sqrt{\frac{2M}{r}},
\end{equation}
where the plus (minus) sign corresponds to outgoing (ingoing)
geodesics respectively. Now we incorporate quantum gravity effects
encoded in the presence of the minimal length and maximal momentum
via the generalized uncertainty principle (1) which motivates
modification of the standard dispersion relation. If the GUP is a
fundamental outcome of quantum gravity proposal, it should appear in
the de Broglie relation as follows (with $\hbar=1$)
\begin{equation}
\lambda\simeq\frac{1}{p}\Big(1-\alpha
\ell_{p}p+\alpha^2\ell_{p}^2p^2\Big),
\end{equation}
or equivalently
\begin{equation}
{\cal{E}}\simeq E\Big(1-\alpha \ell_{p}E+\alpha^2\ell_{p}^2E^2\Big).
\end{equation}

With these preliminaries, we consider a massless particle ejected
from a black hole in the form of a massless shell, and take into
consideration the response of background geometry to radiated
quantum of energy $E$ with GUP correction, i.e. ${\cal{E}}$. The
particle moves on the geodesics of a spacetime with central mass
$M-{\cal{E}}$ substituted for $M$. The description of the motion of
particles in the $s$-wave as spherical massless shells in a
dynamical geometry and the analysis of self-gravitating shells in
Hamiltonian gravity have been reported in Refs. \cite{35,36}. If one
assumes that time increases in the direction of the future, then the
metric should be modified due to back-reaction effects. We set the
total ADM mass, $M$, of the spacetime to be fixed but allow the hole
mass to fluctuate and replace $M$ by $M-{\cal{E}}$ both in the
metric and the geodesic equation. Since the characteristic
wavelength of the radiation is always arbitrarily small near the
horizon due to the infinite blue-shift, the wavenumber approaches
infinity. Therefore the WKB approximation is valid in the vicinity
of the horizon. In the WKB approximation, the tunneling probability
for the classically inhibited area as a function of the imaginary
part of the particle action at the stationary phase takes the form
\begin{equation}
\Gamma\sim\exp(-2\textmd{Im}\, {\cal{I}})\approx\exp(-\beta E)\,.
\end{equation}
As we will see later, to first order in $E$ the right hand side of
this expression substitutes the Boltzman factor in the canonical
ensemble characterized by the inverse temperature $\beta$. In the
$s$-wave picture, particles are spherical massless shells traveling
on the radial null geodesics and transfer across the horizon as
outgoing positive-energy particles from $r_{in}$ to $r_{out}$. The
imaginary part of the action is then given by
\begin{equation}
\textmd{Im}\,
 {\cal{I}}=\textmd{Im}\int_{r_{in}}^{r_{out}}p_rdr=\textmd{Im}\int_{r_{in}}^{r_{out}}\int_0^{p_r}dp'_rdr
\end{equation}
As is clear from the GUP expression, the commutation relation
between the radial coordinate and the conjugate momentum should be
modified as follows
\begin{equation}
[r,p_{r}]=i\Big(1-\alpha \ell_{p}p_{r}+\alpha^2
\ell_{p}^2p_{r}^2\Big).
\end{equation}
In the classical limit it is replaced by the following poisson
bracket
\begin{equation}
\{r,p_{r}\}=1-\alpha \ell_{p}p_{r}+\alpha^2 \ell_{p}^2p_{r}^2.
\end{equation}
Now, using the deformed Hamiltonian equation,
\begin{equation}
\dot{r}=\{r,H\}=\{r,p_r\}\frac{dH}{dr}\bigg{|}_{r}\,,
\end{equation}
since the Hamiltonian is $H=M-{\cal{E}}'$, we assume
$p^2\simeq{\cal{E}}'^2$, $p\simeq{\cal{E}}'$ and eliminate the
momentum in the favor of the energy in integral (7)
\begin{eqnarray}
\textmd{Im}\,{\cal{I}}&=&\textmd{Im}\int_{M}^{M-{\cal{E}}}
\int_{r_{in}}^{r_{out}}\frac{(1-\alpha
\ell_p{\cal{E}}'+\alpha^2 \ell_{p}^{2}{\cal{E}}'^2)}{\dot{r}}drdH\nonumber\\
&=&\textmd{Im}\int_0^{{\cal{E}}}\int_{r_{in}}^{r_{out}}\frac{(1-\alpha
\ell_P{\cal{E}}'+\alpha^2\ell_p^2{\cal{E}}'^2)}{1-\sqrt{\frac{2(M-{\cal{E}}')}{r}}}dr(-d{\cal{E}}')\,.
\end{eqnarray}
The $r$ integral can be performed by deforming the contour around
the pole at the horizon, where it lies along the line of integration
and gives $(-\pi i)$ times the residue
\begin{equation}
\textmd{Im}\, {\cal{I}}=\textmd{Im}\int_0^{{\cal{E}}}4(-\pi
i)\Big(1-\alpha
\ell_p{\cal{E}}'+\alpha^2\ell_p^2{\cal{E}}'^2\Big)\Big(M-{\cal{E}}'\Big)(-d{\cal{E}}')\,.
\end{equation}
This allows us to consider the leading-order correction to be just
proportional to $(\alpha^2\ell_{p}^{2})$ for simplicity and without
loss of generality. Now the imaginary part of the action takes the
following form
\begin{eqnarray}
\textmd{Im}\, {\cal{I}}&=&4\pi ME-2\pi E^2\Big(3M\alpha
\ell_{p}+1\Big)+4\pi
E^3\Big(\frac{7}{3}M\alpha^2\ell_{p}^{2}+\frac{4}{3}\alpha
\ell_p\Big)\nonumber\\&-&10\pi\alpha^2\ell_{p}^{2}E^4
+{\cal{O}}\Big(\alpha^2\ell_{p}^4\Big).
\end{eqnarray}
The tunneling rate is therefore
\begin{eqnarray}
\Gamma&\sim&\exp\Big[-8\pi ME+4\pi E^2(3M\alpha \ell_{p}+1)-8\pi
E^3\Big(\frac{7}{3}M\alpha^{2}\ell_{p}^2+\frac{4}{3}\alpha
\ell_p\Big)\nonumber\\&+&20\pi\alpha^{2}\ell_{p}^{2}E^4
+{\cal{O}}\Big(\alpha^2\ell_{p}^4\Big)\Big]=\exp(\Delta S)\,,
\end{eqnarray}
where $\Delta S$ is the difference in black hole entropies before
and after emission \cite{35,36,37}. In the string theory, it is
anticipated that the tunneling rates from excited $D$-branes in the
microcanonical ensemble depends on the final and initial number of
microstates available to the system. In a more precise expression,
it was shown that the emission rates on the high-energy scales
corresponds to differences between the counting of states in the
microcanonical and in the canonical ensembles \cite{36}. The first
term in the exponential gives a thermal, Boltzmannian spectrum.
However, existence of extra terms shows that the radiation is not
completely thermal. We note that the anticipation of the effect of a
natural cutoff as maximal momentum for emitted particle leads extra
terms in comparison to the results of \cite{24}. These extra terms
enhance the non-thermal behavior of radiation in our setup. In other
words, existence of a maximal momentum enhances the non-thermal
character of the radiation. As equation (14) shows, only to first
order of $E$ the tunneling rate results in the Boltzman factor in
the canonical ensemble characterized by the inverse temperature
$\beta$. So, we see that in the presence of quantum gravity effects,
the emission spectrum cannot be strictly thermal. On the other hand,
if we take the  limit of equation (13) in which back reaction can be
neglected ($\frac{E}{M}\ll 1$), we find $$\textmd{Im}\,
{\cal{I}}=4\pi M E-6\pi\alpha\ell_{p}M
E^{2}+\frac{28}{3}\pi\alpha^{2}\ell_{p}^{2}ME^{3}.$$ This result
shows the pure quantum gravity effects in the absence of back
reaction.

We note that with a GUP that admits just a minimal length, the
emission spectrum takes the following form \cite{24}
\begin{equation}
\Gamma\sim\exp\Big[-8\pi ME+4\pi E^{2}-2\pi\alpha
\ell_{p}^{2}E^{3}\Big(\frac{16}{3}M-5E\Big)+{\cal{O}}\Big(\alpha^2\ell_{p}^4\Big)\Big].
\end{equation}
In comparison between this tunneling rate and the one that we
obtained in this paper as equation (14), it is obvious that when we
consider both maximal momentum and minimal length in the GUP
relation, an additional term of $3M\alpha \ell_P$ appears in the
term that is second order in $E$, which depends on the GUP through
($\alpha\ell_p$). This reflects the fact that in the presence of the
maximal momentum, the tunneling rate deviates from ordinary result
more considerably than the case that we consider just the effect of
minimal length. As we know, in the original Parikh-Wilzcek tunneling
framework, these additional terms are not included since they have
not considered quantum gravitational effects. Since the emitted
particle cannot carry an arbitrary amount of momentum, (in
accordance with Doubly Special Relativity which predicts the
existence of maximal momentum), maximal energy of emitted particles
should be of the order of the Planck energy. Also there are other
additional terms depending on the GUP parameter in second order that
can not be neglected once the black hole mass becomes comparable to
the Planck mass.\\
We now illustrate that the emission rates for the different modes of
radiation during the evaporation are mutually correlated from a
statistical viewpoint. It means that, the total tunneling
probabilities of two particles with energies $E_1$ and $E_2$ are not
similar to tunneling probability of a particle with the total energy
$E_1+E_2$, since there are some correlations between them. Utilizing
(14), the emission rate for a first quantum with energy $E_1$, gives
\begin{equation}
\ln\Gamma_{E_1}=-8\pi ME_1+4\pi E_1^2(3M\alpha \ell_p+1)-8\pi
E_1^3\Big(\frac{7}{3}M\alpha^2\ell_P^2+\frac{4}{3}\alpha
\ell_p\Big)+20\pi\alpha^2\ell_p^2E_1^4.
\end{equation}
Similarly, the emission rate for a single quantum $E_2$, takes the
form
$$\ln\Gamma_{E_{2}}=-8\pi(M-E_{1})E_{2}+4\pi
E_{2}^2\Big(3(M-E_{1})\alpha \ell_{p}+1\Big)$$
\begin{equation}-8\pi
E_{2}^3\Big(\frac{7}{3}(M-E_{1})\alpha^{2}\ell_{p}^2+\frac{4}{3}\alpha
\ell_{p}\Big)+20\pi\alpha^{2}\ell_{p}^{2}E_{2}^4.
\end{equation}
Correspondingly, the emission rate for a single quantum with total
energy, $E=E_{1}+E_{2}$, yields
$$\ln\Gamma_{(E_{1}+E_{2})}=-8\pi
M(E_{1}+E_{2})+4\pi(E_{1}+E_{2})^2(3M\alpha \ell_{p}+1)$$
\begin{equation}
-8\pi(E_{1}+E_{2})^3(\frac{7}{3}M\alpha^{2}\ell_{p}^2+\frac{4}{3}\alpha
\ell_p)+20\pi\alpha^{2}\ell_{p}^{2} (E_{1}+E_{2})^4.
\end{equation}
It can be proved that these probabilities are actually correlated.
The non-zero statistical correlation function is
$$\chi\Big(E_{1}+E_{2};E_{1},E_{2}\Big)=8\pi E_{1}E_{2}\Big(1+12M\alpha
\ell_{p}\Big)$$
\begin{equation}-8\pi E_{1}E_{2}^2\Big(10\alpha \ell_{p}+7M\alpha^{2}\ell_{p}^{2}-\frac{7}{3}E_{2}\alpha^{2}\ell_{p}^2\Big)-8\pi
E_{2}E_{1}^2\Big(7M\alpha^{2}\ell_{p}^{2}+4\alpha \ell_p\Big).
\end{equation}
As we see the statistical correlation function has the terms
depending on the $M$ and $\ell_p$. Also in comparison with the
statistical correlation function obtained in Ref. \cite{24}, the
correlation in our case is enhanced due to incorporation of maximal
momentum for emitted particles. In fact, whenever one quantum of
emission is radiated from the surface of the black hole horizon, the
aberrations are created on the Planck scale that influence the
second quantum of emission, and these aberrations cannot be
neglected, particularly once the black hole mass becomes comparable
with the Planck mass. Therefore as is mentioned in Ref. \cite{24},
in this way the form of modifications as back reaction effects by
incorporation of GUP with minimal length and maximal momentum are
sufficient by themselves to recover information. Information leaks
out from the black holes as the \emph{non-thermal} GUP correlations
within the Hawking radiation.

\section{ Thermodynamics }\vspace{2mm}

\small{ } In this section we study thermodynamics of an evaporating
black hole in the presence of minimal length and maximal momentum.
We mainly focus on the effect of maximal particles' momentum in
thermodynamical quantities.  In the standard viewpoint of black hole
thermodynamics, the temperature and entropy can be obtained easily
by using the standard uncertainty principle \cite{20}. Hawking
considers a vacuum quantum state near the horizon and predicts that
there is a fluctuating sea of virtual pair particles. For instance,
a photon with energy $-E$ is absorbed by the black hole and a photon
with energy $+E$ is emitted to infinity. The energy of the emitted
particle can be obtained by using the standard uncertainty
principle. In fact, in the vicinity of Schwarzschild surface, there
is an intrinsic position uncertainty of the order of Schwarzschild
radius. So, we can write
\begin{equation}
\Delta x\Delta p\sim\hbar \quad\quad \Delta x\simeq
r_s=\frac{2GM}{c^2}\,.
\end{equation}
The uncertainty in the energy of the emitted particle can be
obtained by using the uncertainty in momentum as follows
\begin{equation}
\Delta E=c\Delta p\cong\frac{\hbar c^3}{2GM}
\end{equation}
This is the energy of the emitted particle. With calibration factor
$4\pi$ ,and setting $k_B=1$, the temperature takes the following
form
\begin{equation}
T_H=\frac{c\Delta p}{4\pi}=\frac{\hbar c^3}{8\pi GM}
\end{equation}
As we see, in this approach there is no limit on lowering black hole
mass in the final stage of evaporation. So, in the framework of
Heisenberg uncertainty principle the black hole evaporates totally
in the final stage of its lifetime. According to (22), when the mass
of black hole approaches zero, its temperature becomes infinite. The
entropy of black hole in this case is given by

\begin{equation}
S_B=\frac{4\pi\ G M^{2}}{\hbar c}=\frac{A}{4\ell_{p}^2}\,.
\end{equation}

Now we extend this argument to the case that quantum gravity effects
which are encoded in the GUP are taken into account. Starting with
\begin{equation}
\Delta x\Delta p\geq\hbar\Big(1-\alpha \ell_p\Delta
p+\alpha^2\ell_p^2(\Delta P)^2\Big)
\end{equation}
we solve this equation for the momentum uncertainty in terms of
position uncertainty, which we again takes to be the Schwarzschild
radius $r_s$. This gives the following momentum and temperature for
radiated photons in the presence of maximal particle's momentum and
minimal length
 \begin{equation}
\Delta p=\bigg(\frac{\frac{2M}{M_p^2c}+\alpha
\ell_p}{2\alpha^2\ell_p^2}\bigg)\bigg(1\pm\sqrt{1-\frac{4\alpha^2\ell_p^2}{(\frac{2M}{M_p^2c}+\alpha
\ell_p)^2}}\bigg)
\end{equation}
where we have used $G=\frac{\hbar c}{M_{p}^{2}}$. So we find
\begin{equation}
 T_{(GUP)}=\frac{c\Delta
p}{4\pi}=\bigg(\frac{\frac{2M}{M_p^2}+\alpha
\ell_{p}c}{8\pi\alpha^2\ell_p^2}\bigg)\bigg(1\pm\sqrt{1-\frac{4\alpha^2\ell_p^2}{(\frac{2M}{M_p^2c}+\alpha
\ell_p)^2}}\bigg)\,,
\end{equation}
where we have again used the calibration factor $4\pi$. This result
agrees with the standard result for large masses if the negative
sign is chosen. Whereas the positive sign has no evident physical
meaning. This temperature as a function of mass is shown in Fig.
$1$.

\begin{figure}[htp]
\begin{center}
\includegraphics{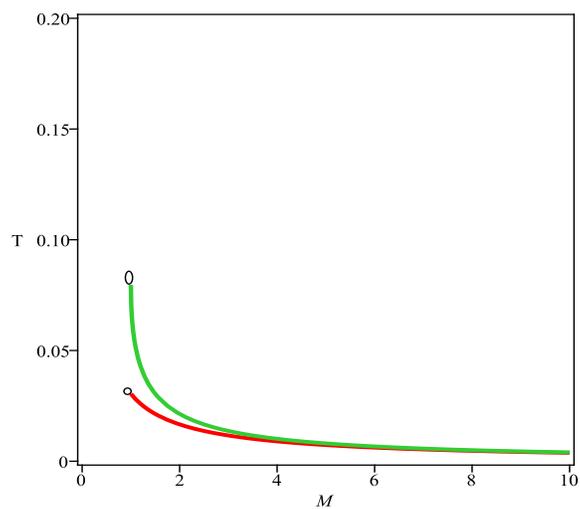}
\end{center}
 \vspace{8cm} \caption{\scriptsize{Temperature of black hole versus its mass.
Mass is in units of the Planck mass and temperature is in units of
Planck energy. The upper curve is temperature that is obtained by
using a GUP admitting just a minimal length and the lower one is
obtained by using a GUP that admits both a minimal length and a
maximal momentum. }}
\end{figure}

The entropy of black hole is obtained by integration of
$dS=\frac{c^2dM}{T}$. In this way we obtain the following
GUP-corrected entropy
\begin{eqnarray}
S_{(GUP)}&=&\frac{2\pi M^2}{M_{p}^2}+2\pi\alpha
\ell_{p}M+\pi\sqrt{\frac{4M^2}{M_{p}^4}+\frac{4M\alpha
\ell_p}{M_{p}^2}
-3\alpha^2\ell_{p}^{2}M}\nonumber\\
&+&8\pi\alpha
\ell_{p}^{2}M_{p}^{2}\sqrt{\frac{4M^{2}}{M_{p}^4}+\frac{4M\alpha
\ell_p}{M_{p}^2}-3\alpha^2\ell_{p}^2}\nonumber\\
&-&2\pi M_{p}^2\alpha^2
\ell_{p}^{2}\ln\bigg[\frac{M_{p}^2}{2}(\frac{2\alpha
\ell_p}{M_{p}^2}+\frac{4M}{M_{p}^4})+\sqrt{\frac{4M^2}{M_{p}^4}+\frac{4M\alpha
\ell_p}{M_{p}^2}-3\alpha^2 \ell_{p}^2}\bigg]
\end{eqnarray}

Note that we have normalized the modified entropy to zero at $M_p$
as shown in Fig. (2).

\begin{figure}[htp]
\begin{center}
\includegraphics{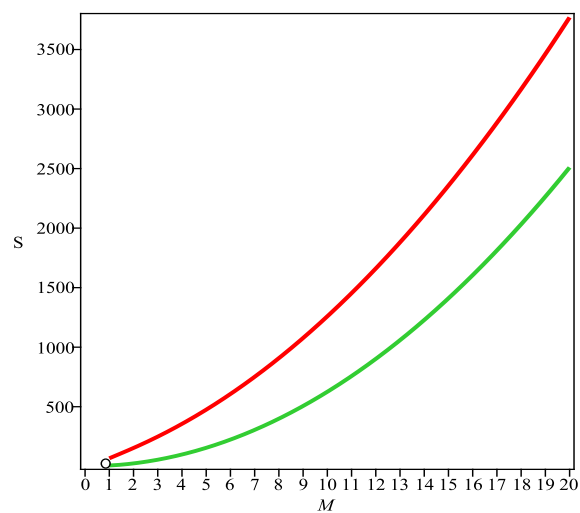}
\end{center}
 \vspace{8cm} \caption{\scriptsize{Entropy of black hole versus its mass. Mass is in the units of the
Planck mass and entropy is in units of the Planck energy. The lower
curve is obtained by using a GUP admitting just a minimal length but
the upper curve is obtained by using a GUP that admits both a
minimal length and a maximal momentum. }}
\end{figure}

This normalization is viable since GUP prevents total evaporation of
black hole (see for instance \cite{21,22}). In this picture, by
incorporating maximal momentum and minimal length, we obtained a
modified temperature and entropy. In fact in this approach, a black
hole should radiate until it approaches the Planck mass and size. At
the Planck mass it ceases to radiate and its entropy vanishes, while
its effective temperature reaches a maximum value. In this phase it
cannot radiate further and becomes an inert \emph{remanent}. We note
that the maximum temperature at the final stage of evaporation when
we consider both maximal momentum and minimal length is smaller than
the case that we consider just the minimal length hypothesis. We
note also that as usual the leading order correction term to the
black hole entropy has a logarithmic nature which is consistent with
loop quantum gravity considerations (see for instance \cite{35,36}).

\newpage

\section{A more general framework}
In Ref. \cite{31} the authors have shown that particles' momentum
cannot be zero if curvature effects are taken into account. In fact,
minimal length and minimal momentum are respectively ultra-violet
and infra-red cutoffs of quantum field theory with gravitational
effect. We note that a framework with a finite minimal uncertainty
$\Delta x_{0}$ can be understood to describe effectively
nonpointlike particles in a fuzzy space. In the same manner, on
large scales a minimal uncertainty $\Delta p_{0}$ may offer new
possibilities to describe situations where momentum cannot be
precisely determined, in particular on curved space. Using the path
integral formulation it has been shown in \cite{31} that such
noncommutative background geometries can ultraviolet and infrared
regularise quantum field theories in arbitrary dimensions through
minimal uncertainties $\Delta x_{0}$ and $\Delta p_{0}$ (for more
details, see \cite{31}. Taking these facts into account, now we can
write a GUP that admits a minimal length, a minimal momentum and a
maximal momentum. This form of GUP is the most general form that
contains all possible natural cutoffs. This GUP can be represented
as (see \cite{31,32,33,34})
\begin{equation}
\Delta x\Delta p\geq\hbar\bigg(1-\alpha\ell_p\Delta
p+\alpha^2\ell_P^2(\Delta p)^2+\beta^2\ell_p^2(\Delta x)^2\bigg)
\end{equation}
This relation has a minimal length which is $(\Delta
x)_{min}=\alpha\ell_P$, a minimal momentum that is $(\Delta
p)_{min}=2\beta\ell_p$ and also a maximal momentum as $(\Delta
p)_{max}=\frac{1}{\alpha
\ell_{p}}\Big(1+\sqrt{1-(1+\beta^{2}\alpha^{2}\ell_{p}^{4})}\Big)$.

Now we want to use this more general GUP to find the tunneling rate
of emitted particles through quantum horizon of black hole like the
way we used in previous sections. By using the Parikh-Wilczek
quantum tunneling mechanism, and quantum gravity effects, we want to
calculate the tunneling rate of emitted particles from Schwarzschild
black hole by incorporation of the above GUP. First, we note that
with this new GUP, the de Broglie relation is represented as
$$\lambda_{\pm}=\frac{p}{2\beta^2\ell_p^2}\bigg(1\pm\sqrt{1-\frac{4\beta^2\ell_p^2(1-\alpha\ell_p p+\alpha^2\ell_p^2
p^2)}{p^2}}\bigg).$$

It is easy to see that positive sign dose not recover ordinary
relation in the limit $\alpha\rightarrow 0$ and $\beta\rightarrow
 0$. So, after expansion of square root we consider the minus sign as
\begin{equation}
\lambda_{-}=\frac{1}{p}\Big(1-\alpha\ell_P p+\alpha^2\ell_P^2
p^2\Big)\bigg(1+\frac{\beta^2\ell_p^2}{p^2}(1-\alpha\ell_P
p+\alpha^2\ell_P^2 p^2)\bigg),
\end{equation}
or equivalently
\begin{equation}
{\cal{E}}=E\Big(1-\alpha\ell_P E+\alpha^2\ell_P^2
E^2\Big)\bigg(1+\frac{\beta^2\ell_P^2}{E^2}(1-\alpha\ell_P
E+\alpha^2\ell_P^2 E^2)\bigg)\,.
\end{equation}
Also as it is clear from the above GUP that the commutation relation
between the radial coordinate and its conjugate momentum should be
modified as

\begin{equation}
[r,p_r]=i\bigg(1-\alpha\ell_p
p+\alpha^2\ell_p^2p^2+\beta^2\ell_p^2r^2\bigg)
\end{equation}
So, in the classical limit, it is replaced by the following poisson
bracket
\begin{equation}
\{r,p_r\}=\bigg(1-\alpha\ell_p
p+\alpha^2\ell_p^2p^2+\beta^2\ell_p^2r^2\bigg).
\end{equation}
With these relations and by using the Parikh-Wilczek method, the
imaginary part of the action takes the following form
\begin{eqnarray}
\textmd{Im}\, {\cal{I}}&=&\frac{4\pi
M\beta^2\ell_p^2}{E}-4\pi\beta^2\ell_p^2-8\pi
M\alpha\beta^2\ell_p^3-4\pi\beta^2\ell_p^2 M^4+4\pi
E\bigg(M-M\alpha\ell_P+4\alpha\beta^2\ell_P^3\nonumber\\&+&2M\alpha^2\beta^2\ell_p^4+4\beta^2\ell_P^2M^3-4\alpha\beta^2\ell_P^3M^3\bigg)-2\pi
E^2\bigg(1-6M\alpha^2\ell_P^2-2\alpha\ell_P\nonumber\\&+&22\alpha^2\beta^2\ell_P^4+\alpha^2\ell_P^2+2M\alpha\ell_P-8\alpha^2\beta^2\ell_P^4M^3+12\beta^2\ell_p^2M^2-24\alpha\beta^2\ell_P^3M^3
+12\alpha^2\beta^2\ell_p^4M^2\bigg)\nonumber\\&+&4\pi
E^3\bigg(\frac{1}{3}\alpha^2\ell_p^2M-2\alpha^2\ell_p^2+\frac{1}{3}\alpha\ell_p-12\alpha^2\beta^2\ell_p^4M^2+4\beta^2\ell_p^2M-12\alpha\beta^2\ell_p^3M+12\alpha^2\beta^2\ell_p^4M
\bigg)\nonumber\\&-&2\pi
E^4\bigg(\frac{1}{2}\alpha^2\ell_p^2-24\alpha^2\beta^2\ell_p^4M+2\beta^2\ell_p^2-8\alpha\beta^2\ell_p^3+12\alpha^2\beta^2\ell_p^4\bigg)
\end{eqnarray}
Now the tunneling rate can be calculated by using relation (6). It
is easy to see that by incorporation of these natural cutoffs, there
are more new terms with origin on quantum gravity effects than
previous cases, even up to the first order of $E$. These extra terms
enhance the non-thermal character of the radiation. In comparison
between this tunneling rate and the one that we obtained for
instance in section $2$ and also the tunneling rate which is
calculated in \cite{24}, it is obvious that by considering all
natural cutoffs in GUP relation, many additional terms are appeared
that shows strong deviation from ordinary thermal radiation. Then,
we can calculate the non-zero statistical correlation function
$\chi(E_1+E_2;E_1,E_2)$ to see that there are more correlation terms
in this case and therefore more possibility to recover information
as non-thermal correlations.

Now we study the thermodynamics of an evaporating black hole in the
presence of minimal length, minimal momentum and also maximal
momentum that are encoded in the GUP relation (28). We saturate this
inequality and solve it for momentum uncertainty in terms of
position uncertainty, which we again take to be the Schwarzschild
radius $r_s$. So we find
\begin{equation}
\Delta p=\frac{1}{2\alpha^2\ell_p^2}\bigg((\Delta
x+\alpha\ell_p)\pm\sqrt{(\Delta
x+\alpha\ell_p)^2-4\alpha^{2}\ell_p^{2}(1+\beta^2\ell_p^2(\Delta
x)^{2})}\bigg).
\end{equation}
Since
$$T=\frac{c\Delta p}{4\pi}$$
we find
\begin{equation}
T=\frac{c}{8\pi\alpha^2\ell_p^2}\bigg((\frac{2GM}{c^2}+\alpha\ell_P)\pm\sqrt{(\frac{2GM}{c^2}+\alpha\ell_P)^2-4\alpha^2\ell_p^2(1+\frac{4\beta^2\ell_p^2G^2M^2}{c^4})}\bigg)
\end{equation}
The negative sign agrees with the standard result for large masses.
The entropy of black hole is derived by integration of $dS=\frac{c^2
dM}{T}$, that is,

\begin{eqnarray}
S&=&\frac{\pi}{2}\frac{M^2 G}{c}+\frac{\pi}{2}Mc\alpha\ell_P+\pi
c\sqrt{\frac{4M^2
G^2}{c^4}+\frac{4M\alpha\ell_pG}{c^2}-3M\alpha^2\ell_p^2}
\nonumber\\ &+&\frac{\pi}{2}G\alpha\ell_p\sqrt{\frac{4M^2
G^2}{c^4}+\frac{4M\alpha\ell_pG}{c^2}-3\alpha^2\ell_p^2}\nonumber\\
&-&\frac{\pi}{2}\frac{\alpha^2\ell_p^2c^3}{G}\ln\bigg[(\alpha\ell_p+\frac{2MG}{c^2})+\sqrt{\frac{4M^2
G^2}{c^4}+\frac{4M\alpha\ell_pG}{c^2}-3\alpha^2\ell_p^2}\bigg]
\end{eqnarray}
We have normalized this modified entropy to zero at $M_P$. In this
way, black hole radiates until it approaches the Planck mass. At
this stage it ceases to radiate and its entropy vanishes leading to
an stable remnant. In Fig. 3 we have compared entropies calculated
in three different situations: the upper curve is for the case that
the applied GUP admits a minimal length and a maximal momentum. The
middle curve is obtained by a GUP that admits just a minimal length,
and finally the lower curve contains the effects of all natural
cutoffs as minimal length, minimal momentum and maximal momentum.
Since $\frac{dS}{dM}=\frac{c^2}{T}$, we see from slopes of these
curves that in the presence of all natural cutoffs, temperature of
black hole for a given mass increases in comparison with the cases
that some of these cutoffs are not taken into account.

\begin{figure}[htp]
\begin{center}
\includegraphics{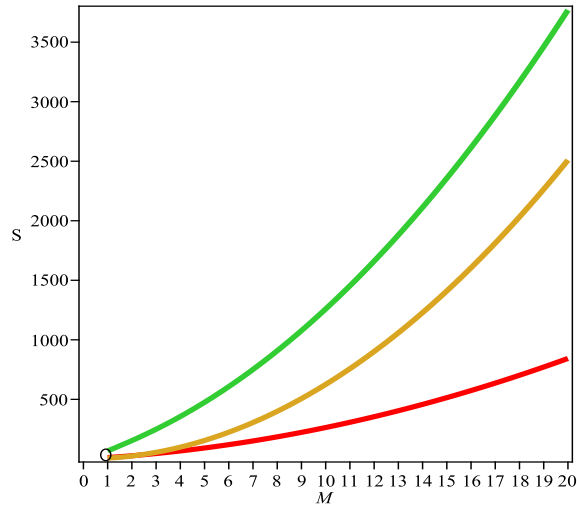}
\end{center}
\vspace{8cm} \caption{\scriptsize{Entropy of black hole versus its
mass for three different GUPs. Mass is in the units of Planck mass
and entropy is in the units of Planck energy. The upper curve is
obtained by using a GUP that admits a minimal length and a maximal
momentum, the middle one with a GUP that admits just a minimal
length and the lower one is obtained with GUP (28) that admits all
natural cutoffs. The slopes of curves reflect the role of natural
cutoffs in each case.}}
\end{figure}

\section{Radiation rate of micro black holes with natural cutoffs}
Any black hole which has a temperature more than the CMB temperature
($2.725$K), should radiate its energy as photons and other ordinary
particles. So, the mass of a black hole decreases and its
temperature increases as its radiates. If we assume that dominant
radiation is photons, we can use the Stefan-Boltzmann law to
estimate the mass output as a function of time which is the emission
rate equation. This emission rate was studied for a Schwarzschild
black hole for the standard case and also in the presence of quantum
gravity effects as a GUP with minimal length in \cite{21,38}. In
this section we calculate emission rate of a black hole in the
presence of both a minimal length and maximal momentum. By using the
Stefan-Boltzmann law and the temperature which was obtained by GUP
relation (1), we have
\begin{equation}
\frac{d}{dt}\Big(\frac{M}{M_p}\Big)=\frac{-16}{t_{ch}}\Big(\frac{M}{M_p}\Big)^6\bigg(1-\sqrt{1-\frac{4\ell_p^2}{\Big(2\ell_p(\frac{M}{M_p})+\ell_p\Big)^{2}}}\bigg).
\end{equation}
where $t_{ch}=60(16)^2\pi T_p$ is a characteristic time and is about
$4.8 \times 10^4$ times the Planck time. The emission rate for a
black hole by using the GUP with just a minimal length is \cite{21}
\begin{equation}
\frac{d}{dt}\Big(\frac{M}{M_p}\Big)=\frac{-16}{t_{ch}}\Big(\frac{M}{M_p}\Big)^6\bigg(1-\sqrt{1-\frac{1}{(\frac{M}{M_p})^2}}\bigg).
\end{equation}
These emission rates as a function of mass are shown in Fig. 4. We
see that at the final stage of evaporation, when the mass of black
hole is of the order of the Planck mass, the emission rate reaches a
maximum as black hole temperature increases. Note also that the
emission rate at the final stage of evaporation with a GUP that
admits just a minimal length is larger than the corresponding
quantity in the presence of both minimal length and maximal
momentum. This feature coincides with the result that we obtained in
section 3, which tells us that temperature of black hole in the
final stage of its evaporation in the presence of both minimal
length and maximal momentum is less than the case that we consider
just the effect of minimal length. On the other hand, the GUP
relation (1) with minimal length and maximal momentum leads to
further reduction of black hole lifetime. Black hole in the presence
of natural cutoffs as minimal length and maximal momentum should
evaporate sooner than the case that we consider just a minimal
length.

\begin{figure}[htp]
\begin{center}
\includegraphics{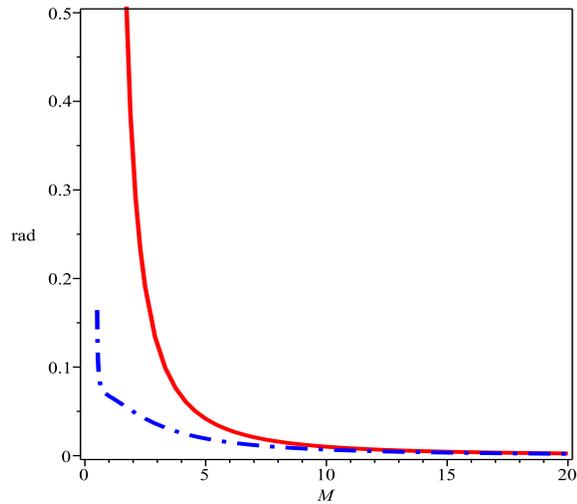}
\end{center}
\vspace{8cm} \caption{\scriptsize{Radiation rate versus the mass.
The mass is in the units of Planck mass. The lower curve is obtained
by relation (51), and the upper one is obtained by relation (52). }}
\end{figure}

\section{Comparison and Conclusion}

In this paper we considered possible effects of natural cutoffs as a
minimal length, a maximal momentum, and a minimal momentum on the
tunneling rate and thermodynamics of TeV scale black holes. In this
respect we considered two different GUPs. While previous studies in
this field have considered just the effect of minimal length and
ignored possible cutoffs on emitted particles' momentum, here we
considered the effects of all possible cutoffs on length and
momentum measurement. We considered two types of GUPs: one with a
minimal length and maximal momentum, and the other with all natural
cutoffs including minimal length, minimal momentum, and maximal
momentum. The later one is the most general form of GUP adopted to
study TeV black hole thermodynamics. We have shown that in the
presence of all natural cutoffs, the tunneling rate is strongly
deviated from thermal emission. Even in the first order of $E$,
extra terms that depend on the GUP parameters are appeared in the
tunneling rate relation, leading to more deviation from thermal
emission in comparison with the case that one considers just the
effect of minimal length. Also we have shown that a part of
information can be recovered as correlations between emitted modes
in the presence of all natural cutoffs. In the final stage of black
hole evaporation, both adopted GUPs predict a minimal mass remnant
for black hole. Part of information can be supplied in this Planck
size remnant. We have shown also that, at the final stage of
evaporation black hole reaches a maximum temperature and its entropy
becomes zero, so it turns into a stable remanent. As both GUPs have
a minimal length $\Delta x_{min}=\alpha\ell_p$, the minimum mass
would be $M_{min}=\frac{\alpha\ell_pc^2}{2G}$. By using this minimum
mass we found the final temperature of black hole in the presence of
both GUPs. With a GUP as given by relation (1) that admits a minimal
length and a maximal momentum, we have
$T_{final}=\frac{c}{4\pi\alpha\ell_p}$. On the other hand, with most
general GUP (28) admitting all natural cutoffs, we obtain
$T_{final}=\frac{c(1+\beta\alpha\ell_p)}{4\pi\alpha\ell_p}$. With
these results in mind, we see that the GUP with minimal length and
maximal momentum predicts that the final state temperature is less
than the case that we consider just a minimal length GUP.
Nevertheless, when we consider all natural cutoffs through GUP (28),
the maximum temperature is larger than the case that one ignores the
existence of minimal momentum or considers just the effect of
minimal length. In fact, the GUP with all natural cutoffs shows that
at the final stage of evaporation of black hole, its temperature
increases as it is clear from the slopes of the curves in Fig 3.
Another important impact of these natural cutoffs on black hole
thermodynamics can be seen in black hole radiation rate.

At the final stage of evaporation, when the mass of black hole is of
the order of the Planck mass, the emission rate reaches a maximum as
black hole temperature increases. The emission rate at the final
stage of evaporation with a GUP that admits just a minimal length is
larger than the corresponding quantity in the presence of both
minimal length and maximal momentum. We have shown also that in the
presence of both minimal length and maximal momentum the lifetime of
black hole decreases relative to the case that we consider just the
effect of minimal length.


\section*{References}
\begin{thebibliography}{100}
\bibitem{1}
G. Veneziano, Europhys. Lett. \textbf{2}, 199 (1986).
\bibitem{2}
D. Amati, M. Cialfaloni and G. Veneziano, Phys. Lett. B \textbf{ 197}, 81 (1987).
\bibitem{3}
D. Amati, M. Cialfaloni and G. Veneziano, Phys. Lett. B \textbf{ 216}, 41 (1989).
\bibitem{4}
D. J. Gross and P. F. Mende, Phys. Lett. B \textbf{ 197}, 129 (1987).
\bibitem{5}
K. Konishi, G. Paffuti and P. Provero, Phys. Lett. B \textbf{ 234}, 276 (1990).
\bibitem{6}
R. Guida, K. Konishi and P. Provero, Mod. Phys. Lett. A \textbf{6}, 1487 (1991).
\bibitem{7}
M. Kato, Phys. Lett. B \textbf{245}, 43 (1990).
\bibitem{8}
L. J. Garay, Int. J. Mod. Phys. A \textbf{10}, 145 (1995).
\bibitem{9}
S. Capozziello, G. Lambiase and G. Scarpetta, Int. J. Theor. Phys. \textbf{39}, 15 (2000).
\bibitem{10}
M. Maggiore, Phys. Lett. B \textbf{304}, 65 (1993).
\bibitem{11}
M. Maggiore, Phys. Lett. B \textbf{319}, 83 (1993).
\bibitem{12}
M. Maggiore, Phys. Rev. \textbf{49}, 5182 (1994).
\bibitem{13}
S. Hossenfelder, [arXiv:1203.6191]; S. Hossenfelder, Class. Quantum Grav. \textbf{29}, 115011 (2012).
\bibitem{14}
F. Scardigli, Phys. Lett. B \textbf{452}, 39 (1999).
\bibitem{15}
G. Amelino-Camelia, Int. J. Mod. Phys. D \textbf{11}, 35
(2000); G. Amelino-Camelia, Nature \textbf{418}, 34 (2002); G.
Amelino-Camelia,  Int. J. Mod. Phys. D \textbf{11}, 1643 (2002); J.
Kowalski-Glikman,  Lect. Notes Phys. \textbf{669}, 131 (2005); G.
Amelino-Camelia, J. Kowalski-Glikman, G. Mandanici and A.
Procaccini,  Int. J. Mod. Phys. A \textbf{20}, 6007 (2005); K.
Imilkowska, J. Kowalski-Glikman, Lect. Notes Phys.\textbf{702}, 279
(2006).
\bibitem{16}
J. Magueijo and L. Smolin, Phys. Rev. Lett. \textbf{88}, 190403 (2002).
\bibitem{17}
J. Magueijo and L. Smolin, Phys. Rev. Lett. D \textbf{ 67}, 044017 (2003).
\bibitem{18}
J. Magueijo and L. Smolin, Phys. Rev. D \textbf{71}, 026010 (2005).
\bibitem{19}
J. L. Cortes and J. Gamboa, Phys. Rev. D \textbf{71}, 065015 (2005).
\bibitem{20}
H. Ohanian and R. Ruffini, \emph{Gravitation and
Spacetime}, 2nd edition (W. W. Norton) 1994, p. 481.
\bibitem{21}
R. J. Adler, P. Chen and D. I. Santiago, Gen. Rel. Grav.
\textbf{33}, 2101 (2001).
\bibitem{22}
G. Amelino-Camelia, M. Arzano, Y. Ling and G. Mandanici,
Class. Quant. Grav. \textbf{23}, 2585 (2006).
\bibitem{23}
K. Nozari and A. S. Sefiedgar, Phys. Lett. B \textbf{635}, 156
(2006); K. Nozari and A. S. Sefiedgar, Gen. Rel. Grav. \textbf{39},
501 (2007).
\bibitem{24}
K. Nozari and S. H. Mehdipour, Europhys. Lett. \textbf{84}, 20008
(2008).
\bibitem{25}
K. Nozari and S. H. Mehdipour, Class. Quantum Grav. \textbf{25},
175015 (2008); K. Nozari and S. H. Mehdipour, JHEP \textbf{0903},
061 (2009).
\bibitem{26}
W. Kim, E. J. Son and M. Yoon, JHEP \textbf{0801}, 035
(2008).
\bibitem{27}
L. Xiang and X. Q. Wen, JHEP \textbf{0910}, 046 (2009).
\bibitem{28}
R. Banerjee and S. Ghosh, Phys. Lett. B, \textbf{688}, 224
(2010).
\bibitem{29}
M. K. Parikh and F. Wilczek, Phys. Rev. Lett.,
\textbf{85}, 5042 (2000).
\bibitem{30}
K. Nozari and A. Etemadi, Phys. Rev. D \textbf{85}, 104029 (2012).
\bibitem{31}
H. Hinrichsen and A. Kempf, J. Math. Phys. \textbf{37}, 2121 (1996);
A. Kempf, J. Math. Phys. \textbf{38}, 1347 (1997); A. Kempf,
Preprint DAMTP/94-33, hep-th/9405067 (1994).
\bibitem{32}
A. F. Ali, S. Das and E.C. Vagenas, Phys. Lett. B \textbf{678}, 497
(2009); S. Das, E. C. Vagenas and A. F. Ali, Phys. Lett. B
\textbf{690}, 407 (2010).
\bibitem{33}
P. Pedram, K. Nozari and S. H. Taheri, JHEP \textbf{03}, 093 (2011).
\bibitem{34}
A. Kempf, G. Mangano and R. B. Mann, Phys. Rev. D \textbf{52}, 1108
(1995).
\bibitem{35}
P. Kraus and F. Wilczek, Nucl. Phys. B \textbf{433}, 403
(1995).
\bibitem{36}
E. C. Vagenas, Phys. Lett. B \textbf{503}, 399 (2001);
E. C. Vagenas, Mod. Phys. Lett. A \textbf{17}, 609 (2002); E. C.
Vagenas, Phys. Lett. B \textbf{533}, 302 (2002);  A. J. M. Medved,
Class. Quantum Grav. \textbf{19}, 589 (2002); A. J. M. Medved, Phys.
Rev. D \textbf{66},  124009 (2002); E. C. Vagenas, Phys. Lett. B
\textbf{559}, 65 (2003); M. K. Parikh, Int. J. Mod. Phys. D
\textbf{13}, 2351 (2004); E. T. Akhmedov, V. Akhmedova and D.
Singleton, Phys. Lett. B \textbf{642}, 124 (2006); E. T. Akhmedov,
V. Akhmedova, T. Pilling and D. Singleton, Int. J. Mod. Phys. A
\textbf{22}, 1705 (2007).
\bibitem{37}
E. Keski-Vakkuri and P. Kraus, Nucl. Phys. B
\textbf{491}, (249) (1997).
\bibitem{38}
M. Cavaglia and S. Das, Class. Quantum Grav. \textbf{21}, 4511
(2004). See also: V. Cardoso, M. Cavaglia and L. Gualtieri, Phys.
Rev. Lett. \textbf{96}, 071301 (2006); Erratum-ibid. \textbf{96},
219902 (2006); V. Cardoso, M. Cavaglia and L. Gualtieri, JHEP
\textbf{0602}, 021 (2006).




\end{thebibliography}
\end{document}